# On the construction of bivariate linear exponential distribution with FGM family

**M. A. El-Damcese[1], Dina. A. Ramadan[2]**

[1] Mathematics Department, Faculty of Science, Tanta University, Tanta, Egypt

(e-mail:meldamcese@yahoo.com)

[2] Mathematics Department, Faculty of Science, Mansoura University, Mansoura,, Egypt

## Abstract

In this paper we propose a Farlie-Gumbel-Morgenstern (*FGM*) family of bivariate linear exponential distributions generated from given marginal's. Therefore, properties of FGM are analogous to properties of bivariate distributions. We study some important statistical properties and results for the new distribution.

## 1. Introduction

In 1960 Gumbel investigated the same structure for exponential marginal. Also, in 1960, **Farlie** — in connection with his investigations of the correlation coefficient — suggested a generalization of the bivariate form studied by **Morgenstern** and **Gumbel**. **Johnson** and **Kotz** (1975) (1977) studied the multivariate case and coined the term the *FGM* distribution. Further studies were carried out by **Schucany et al.** (1978), **Kotz** and **Johnson** (1977), and **Huang** and **Kotz** (1984), inter alia. The last two papers deal with an "iterated"Farlie-Gumbel-Morgenstern distribution. **Shaked** (1975) investigates the relation between the *FGM* distribution family and the so- called "positive dependent in mixture" distributions. **Johnson** (1980) utilizes the *FGM* family for models involving censoring, **D'este** (1981) analyzes the structure of the *FGM* distribution with gamma marginal, **Lin** (1987) discusses various parameterizations of the *FGM* family. **Nelson** (1994) characterizes the *FGM* family in terms of the value of the correlation coefficient between the marginals. **De La Horra** and **Fernandez** (1995) propose the *FGM* family as a class of robust prior distributions. **Huang** and **Kotz** (1999) developed Polynomial-type single parameter extensions of the Farlie-Gumbel-Morgenstern bivariate distributions.

*Bairamov* and *Kotz* (2002) obtained the some theorems characterizing symmetry for Huang-Kotz *FGM* distributions and conditions for independence. *Lai* and *Xie* (2000), *Drouet-Mari* and *Kotz* (2001) investigated the relationship between the *FGM* distribution family and the so-called "positive dependent in mixture" distributions. *Kim* and *Sungur* (2004) utilized the *FGM* family to models involved censoring. *Durante* (2006) analyzed the structure of the *FGM* distribution with gamma marginals and discussed the various parameterizations of the *FGM* family. *Durante* and *Jaworski* (2009) derivatives a new characterization of bivariate copula, that is given by using the notion of Dini derivatives. *Kim et al*. (2011) proposed a new class of bivariate copula to quantify dependence and incorporate into various iterated copula families. *Bekrizadeh et al*. (2012) extended the domain of correlation Farlie–Gumbel–Morgenstern copulas and also use it to model high negative dependence values and the ranges of the Spearman's correlation in our proposed extension have been found to be in [-0.5,0.43]. *Carles*, *Cuadras* and *WalterDǏaz* (2012) proposed another generalization for *FGM* and study its properties, after defining the dimension of a distribution as the cardinal of the set of canonical correlations, proved that some well-known distributions are practically two-dimensional and introduced an extended *FGM* family in two dimensions and study how to approximate any distribution to this family.

The paper is organized as follows. In Section 2, we derive model of the linear exponential distributions. Expressions for the reliability and MTTF of the (*FGM*) family of bivariate linear exponential distribution are presented in the Section 3. The two-dimensional failure modeling and its minimal repair and replacement discussed in the Section 4. An expression for monotonicity of the (reversed) hazard rate is given in the Section 5. Properties of bivariate *FGM* bivariate linear exponential distribution with proportional hazard rate models are discussed in the Section6. Finally, we conclude the paper in the Section 7.

## 2. Model

The bivariate Farlie Gumbel Morgenstern (*FGM*) distribution originally introduced by *Morgenstern* (1956) describes a system consisting of two dependent components. Various tests of independence are a statistical tool used to investigate the dependence structure between the components.

The Farlie-Gumbel-Morgenstern distributions has joint cumulative distribution functions of the form

$$F(x,y) = F_X(x)F_Y(y)\left[1+\lambda(1-F_X(x))(1-F_Y(y))\right]$$

$$= \left(1 - e^{-(\alpha_1 x + \frac{1}{2}\beta_1 x^2)}\right)\left(1 - e^{-(\alpha_2 y + \frac{1}{2}\beta_2 y^2)}\right)\left[1 + \lambda e^{-(\alpha_1 x + \alpha_2 y + \frac{1}{2}(\beta_1 x^2 + \beta_2 y^2))}\right]$$

$$|\lambda| \leq 1, x, y \geq 0 \qquad (1)$$

where $F_X$ and $F_Y$ are the marginal cumulative.

A Joint *FGM* cumulative distribution following from equation (1) is illustrated in Figure 1 for several dependence parameter values $\lambda$ and with parameter values $\alpha_1 = 0.5, \alpha_2 = 0.7, \beta_1 = 1.5$ and $\beta_2 = 2$.

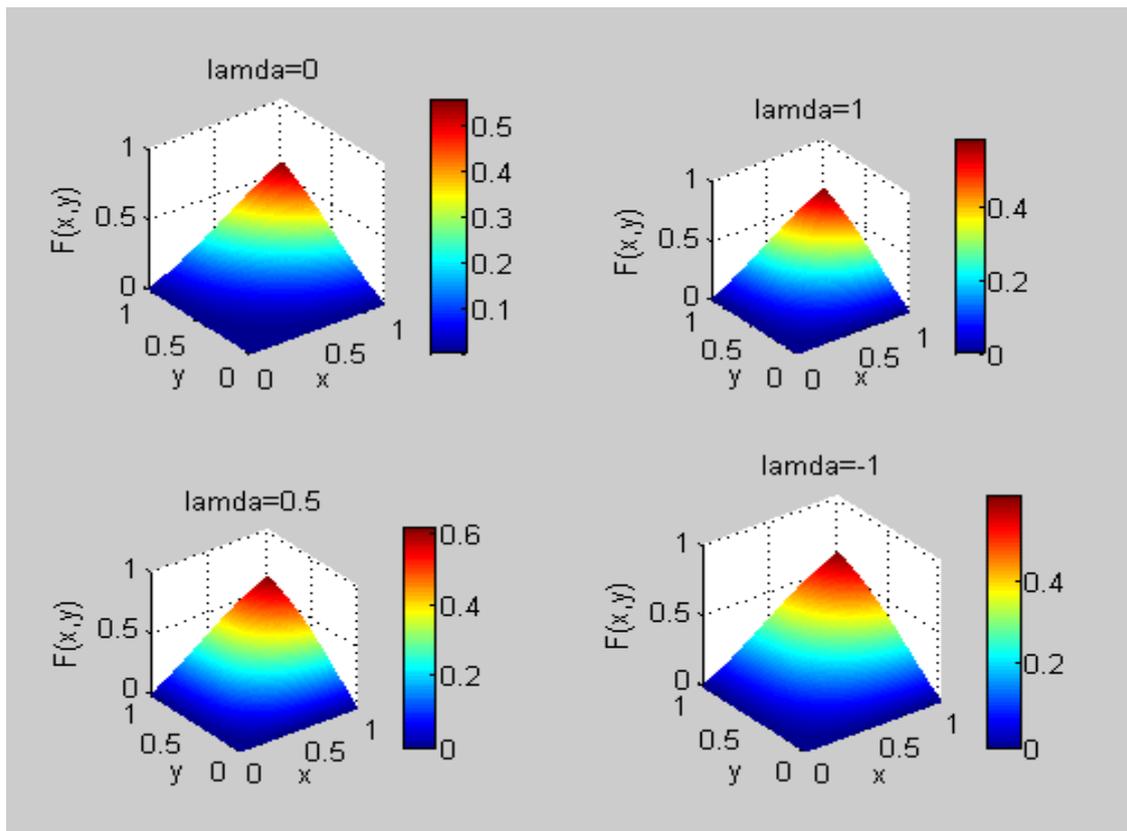

***Figure1.*** *Joint FGM Cumulative Distribution for Different Dependence Values of λ*

The joint density corresponding to (1) is

$$f(x,y) = f_X(x)f_Y(y)[1 + \lambda(2F_X(x) - 1)(2F_Y(y)-1)]$$

$$= (\alpha_1 + \beta_1 x)(\alpha_2 + \beta_2 y)e^{-(\alpha_1 x+\alpha_2 y+\frac{1}{2}(\beta_1 x^2+\beta_2 y^2))}\Big[1 + \lambda\left(1 - 2e^{-(\alpha_1 x+\frac{1}{2}\beta_1 x^2)}\right)\left(1 - 2e^{-(\alpha_2 y+\frac{1}{2}\beta_2 y^2)}\right)\Big]$$

(2)

Where

$$f_X(x) = (\alpha_1 + \beta_1 x)e^{-(\alpha_1 x+\frac{1}{2}\beta_1 x^2)},$$

$$F_X(x) = 1 - e^{-(\alpha_1 x+\frac{1}{2}\beta_1 x^2)},$$

$$f_Y(x) = (\alpha_2 + \beta_2 y)e^{-(\alpha_2 y+\frac{1}{2}\beta_2 y^2)},$$

$$f_Y(x) = 1 - e^{-(\alpha_2 y+\frac{1}{2}\beta_2 y^2)}.$$

Furthermore, the conditional density function and conditional distribution function of X given Y = y, respectively, are

$$f_{Y=y}(x\backslash y) = f_X(x)[1 + \lambda(2F_X(x) - 1)(2F_Y(y)-1)]$$

$$= (\alpha_1 + \beta_1 x)e^{-(\alpha_1 x+\frac{1}{2}\beta_1 x^2)}\Big[1 + \lambda\left(1 - 2e^{-(\alpha_1 x+\frac{1}{2}\beta_1 x^2)}\right)\left(1 - 2e^{-(\alpha_2 y+\frac{1}{2}\beta_2 y^2)}\right)\Big],$$

(3)

$$F_{Y=y}(x\backslash y) = F_X(x)[1 + \lambda(1-F_X(x))(1-F_Y(y))]$$

$$= (1 - e^{-(\alpha_1 x+\frac{1}{2}\beta_1 x^2)})\Big[1 + \lambda e^{-(\alpha_1 x+\alpha_2 y+\frac{1}{2}(\beta_1 x^2+\beta_2 y^2))}\Big].$$

(4)

The correlation coefficient is

$$\rho = corr\{F_X(x), F_Y(y)\} = corr\left\{1 - e^{-(\alpha_1 x+\frac{1}{2}\beta_1 x^2)}, 1 - e^{-(\alpha_2 y+\frac{1}{2}\beta_2 y^2)}\right\}.$$

(5)

## 3. Bivariate Reliability Models

There are certainly many ways to define a bivariate reliability model. In addition, there are probably several alternate ways to classify the model types. We feel that aninformative classification scheme is based on the relationship between the two variables.Specifically, we distinguish between those models for which age and use are functionallyrelated and those in which they are correlated rather than functionally dependent.

We further separate the models in which the two variables are functionally related onthe basis of whether the functions are deterministic or stochastic. The models based oncorrelation of the two variables may be further classified by whether $\rho = 0$ or $\rho \neq 0$. Ofcourse, from a reliability perspective, the case in which age and use are independent is$\rho$ unlikely to be practically meaningful.

The reliability function for *FGM* is

$$\bar{F}(x,y) = P[X \geq x, Y \geq y] = \int_x^\infty \int_y^\infty f(u,v) du\, dv,$$

(6)

so,

$$\bar{F}(x,y) = \int_x^\infty \int_y^\infty f_U(u) f_V(v)[1 + \lambda(2F_U(u) - 1)(2F_V(v)-1)]\, dv\, du$$

$$= (1 - F_X(x))(1 - F_Y(y))[1 + \lambda F_X(x) F_Y(y)],$$

$$= e^{-(\alpha_1 x + \alpha_2 y + \frac{1}{2}(\beta_1 x^2 + \beta_2 y^2))} \left[1 + \lambda \left(1 - e^{-(\alpha_1 x + \frac{1}{2}\beta_1 x^2)}\right)\left(1 - e^{-(\alpha_2 y + \frac{1}{2}\beta_2 y^2)}\right)\right].$$

(7)

Reliability Function for *FGM* following from equation (7) is illustrated in Figure 2 for several dependence parameter values $\lambda$ and with parameter values $\alpha_1 = 0.05, \alpha_2 = 0.07, \beta_1 = .15$ and $\beta_2 = .2$.

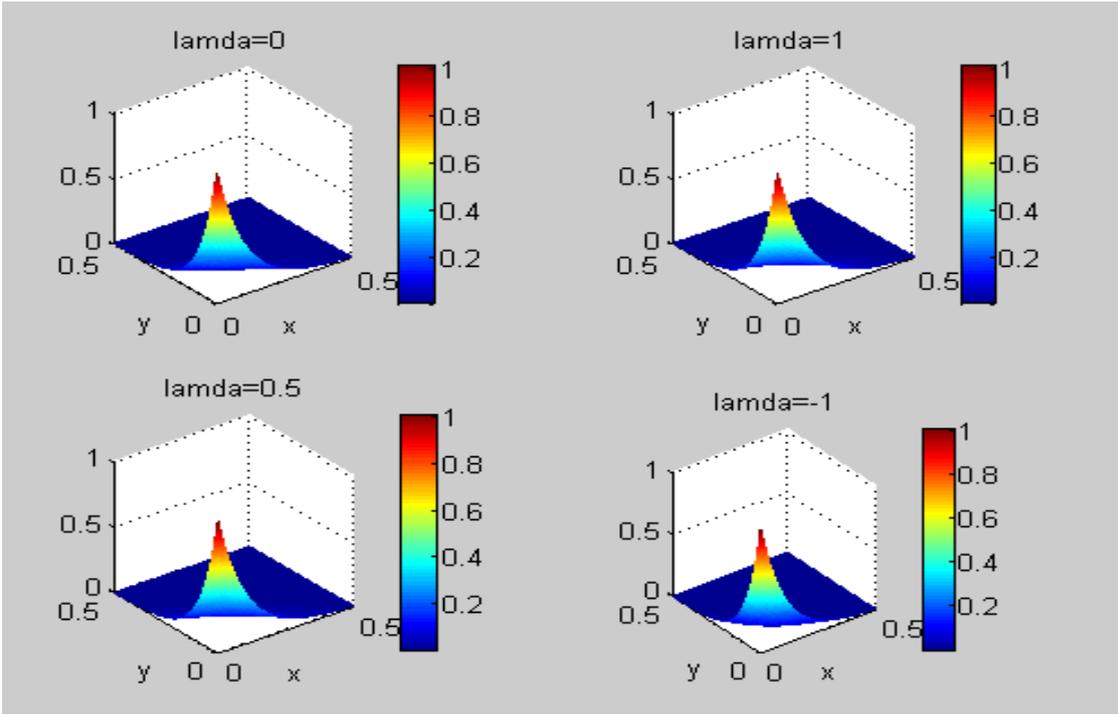

***Figur2.** Reliability Function for FGM with Different Dependence Values of λ*

Usually we are interested in the expected time to next failure, and this is termed mean time to failure. The mean time to failure (*MTTF*) is defined as the expected value of the lifetime before a failure occurs.

Suppose that the reliability function for *FGM* distribution is given by $\bar{F}(x,y)$, the *MTTF* can be computed as

$$MTTF = \int_0^\infty \int_0^\infty x\,y\,f(x,y)dy\,dx = \int_0^\infty \int_0^\infty \bar{F}(x,y)dy\,dx$$

(8)

Then,

$$MTTF = \int_0^\infty \int_0^\infty (1 - F_X(x))(1 - F_Y(y))[1 + \lambda F_X(x)F_Y(y)]dy\,dx$$

$$= \int_0^\infty (1 - F_X(x)) \int_0^\infty [(1 - F_Y(y)) + \lambda F_X(x)F_Y(y)(1 - F_Y(y))]dy\,dx,$$

(10)

The series expansions of $e^{-\frac{1}{2}\beta_1 x^2}$ and $e^{-\frac{1}{2}\beta_2 y^2}$ are

$$e^{-\frac{1}{2}\beta_1 x^2} = \sum_{m=0}^{\infty} \frac{(-1)^m \beta_1^m x^{2m}}{2^m \, m!},$$

(11)

and

$$e^{-\frac{1}{2}\beta_2 y^2} = \sum_{n=0}^{\infty} \frac{(-1)^n \beta_2^n y^{2n}}{2^n \, n!}.$$

(12)

We get,

$$\int_0^{\infty} (1 - F_X(x))\, dx = \sum_{m=0}^{\infty} \frac{(-1)^m \beta_1^m \, \Gamma(2m+1)}{2^m \, \alpha_1^{2m} \, m!},$$

(13)

$$\int_0^{\infty} F_X(x)(1 - F_X(x))\, dx = \sum_{m=0}^{\infty} \frac{(-1)^m \beta_1^m \, \Gamma(2m+1)}{2^m \, \alpha_1^{2m} \, m!} \left(\frac{2^m - 1}{2^m}\right),$$

(14)

$$\int_0^{\infty} (1 - F_Y(y))\, dy = \sum_{n=0}^{\infty} \frac{(-1)^n \beta_2^n \, \Gamma(2n+1)}{2^n \, \alpha_2^{2n} \, n!},$$

(15)

and

$$\int_0^{\infty} F_Y(y)(1 - F_Y(y))\, dy = \sum_{n=0}^{\infty} \frac{(-1)^n \beta_2^n \, \Gamma(2n+1)}{2^n \, \alpha_2^{2n} \, n!} \left(\frac{2^n - 1}{2^n}\right),$$

(16)

Substituting (13), (14), (15) and (16) into (10), we obtained

$$MTTF = \sum_{m=n=0}^{\infty} \frac{(-1)^{m+n} \beta_1^m \beta_2^n \Gamma(2m+1)\, \Gamma(2n+1)}{2^{m+n} \, \alpha_1^{2m} \alpha_2^{2n} \, m! \, n!} \left[1 + \lambda \, \frac{(2^m - 1)(2^n - 1)}{2^{m+n}}\right].$$

## 4. Two-Dimensional Failure Modeling

We now assume that the degradation of a system depends on its age and usage. Let $X_n$ and $Y_n$, $n > 1$, denote the time of the $n^{th}$ system failure and the corresponding usage at that time. $Z_n = X_n - X_{n-1}$ gives the time between the $n^{th}$ and $(n-1)^{th}$ failure, and $T_n = Y_n - Y_{n-1}$ is the system usage during this period (where $X_0 = 0, Y_0 = 0$).

The hazard function is defined as

$$r(x,y) = \frac{f(x,y)}{\bar{F}(x,y)},$$

then,

$$r(x,y) = \frac{(\alpha_1 + \beta_1 x)(\alpha_2 + \beta_2 y)\left[1 + \lambda\left(1 - 2e^{-(\alpha_1 x + \frac{1}{2}\beta_1 x^2)}\right)\left(1 - 2e^{-(\alpha_2 y + \frac{1}{2}\beta_2 y^2)}\right)\right]}{\left[1 + \lambda\left(1 - e^{-(\alpha_1 x + \frac{1}{2}\beta_1 x^2)}\right)\left(1 - e^{-(\alpha_2 y + \frac{1}{2}\beta_2 y^2)}\right)\right]}$$

(18)

So the probability that the first system failure will occur in $[x, x + \delta x) \times [y, y + \delta y)$ given that $X_1 > x$ and $Y_1 > y$ is $\lambda(x,y) \delta x \delta y + o(\delta x \delta y)$.

Successive system failures can be modeled using a two-dimensional point process formulation. We let $N_2(s, x; w, y)$ denote the number of system failures in the rectangle $[s, x) \times [w, y)$, $0 \leq s < x$, $0 \leq w < y$ and we abbreviate $N_2(0, x; 0, y)$ to $N_2(x, y)$. The failure intensity or rate of occurrence of failures (**ROCOF**) at the point (x, y) is given by the function

$$\lambda(x,y) = \lim_{\substack{\delta x \to 0 \\ \delta y \to 0}} \frac{P[N_2(x, x+\delta x; y, y+\delta y) = 1]}{\delta x \delta y}$$

(19)

so the probability that a failure will occur in $[x, x + \delta x) \times [y, y + \delta y)$ is $\lambda(x,y) \delta x \delta y + o(\delta x \delta y)$.

Assuming simultaneous failures cannot occur, it follows from (19) that $\lambda(x,y) = \frac{\partial^2 E[N_2(x,y)]}{\partial x\, \partial y}$ or

$$E[N_2(x,y)] = \Lambda(x,y) = \int_0^x \int_0^y r(u,v)\, dv\, du$$

(20)

Let $H_{x,y}$ denote the history of the failure process up to, but not including, the point (x, y). The conditional failure intensity function is then given by

$$\lambda_c(x,y) = \lim_{\substack{\delta x \to 0 \\ \delta y \to 0}} \frac{P[N_2(x, x+\delta x;\, y, y+\delta y) = 1 / H_{x,y}]}{\delta x\, \delta y}$$

(21)

The forms for the functions $\lambda_c(x,y)$, $\lambda(x,y)$ and $N_2(x,y)$ are now discussed for the two cases "always minimally repair" and "always replace".

## 4.1 Minimal Repair

The conditional failure intensity function for *FGM* is unaffected by each failure and so

$$\lambda_c(x,y) = \lambda(x,y) = r(x,y),$$

(22)

where $r(x,y)$ is the hazard function for $(X_1, Y_1)$ given in (18). We now show that the counting process $\{N_2(x,y), x \geq 0, y \geq 0\}$ is a two-dimensional *NHPP* with intensity function given by (21), so that

$$E[N_2(x,y)] = \Lambda(x,y) = \int_0^x \int_0^y r(u,v)\, dv\, du$$

(23)

*Proof*

$$\Lambda(x,y) = \int_0^x \int_0^y \frac{f_U(u)f_V(v)[1 + \lambda(2F_U(u) - 1)(2F_V(v)-1)]}{(1 - F_U(u))(1 - F_V(v))[1 + \lambda F_U(u)F_V(v)]} \, dv \, du$$

(24)

Since $0 < e^{-(\alpha_1 x + \frac{1}{2}\beta_1 x^2)} < 1$ and $0 < e^{-(\alpha_2 y + \frac{1}{2}\beta_2 y^2)} < 1$ for $x$ and $y$, then by using the binomial series expansion

$$[(1 - F_U(u))(1 - F_V(v)) + \lambda F_U(u)F_V(v)(1 - F_U(u))(1 - F_V(v))]^{-1}$$
$$= \sum_{k=0}^{\infty} (-1)^k \lambda^k (1 - F_U(u))^{-(k+1)} (1 - F_V(v))^{-(k+1)} \left(F_U(u) - (F_U(u))^2\right)^k \left(F_V(v) - (F_V(v))^2\right)^k,$$

(25)

$$(1 - F_U(u))^{-(k+1)} = \sum_{r=0}^{k} \binom{k+r}{r} (F_U(u))^r,$$

(26)

$$(1 - F_V(v))^{-(k+1)} = \sum_{s=0}^{k} \binom{k+s}{s} (F_V(v))^s,$$

(27)

$$\left(F_U(u) - (F_U(u))^2\right)^k = \sum_{n=0}^{\infty} (-1)^n \binom{k}{n} (F_U(u))^{k+n}$$

(28)

and

$$\left(F_V(v) - (F_V(v))^2\right)^k = \sum_{m=0}^{\infty} (-1)^m \binom{k}{m} (F_V(v))^{k+m}$$

(29)

Substituting (25),(26),(27),(28) and (29) into (24), we get

$\Lambda(x, y)$

$$= \sum_{k=r=s=n=m=0}^{\infty} (-1)^{k+n+m} (\lambda)^k \binom{k+r}{r} \binom{k+s}{s} \binom{k}{n} \binom{k}{m} \int_0^x f_U(u)(F_U(u))^{k+n+r} \left\{ \int_0^y f_V(v)(F_V(v))^{k+m+s} [1 \right.$$

$$\left. + \lambda(2F_U(u) - 1)(2F_V(v)-1)] \right\} dv\, du$$

Let,

$$C = (-1)^{k+n+m} (\lambda)^k \binom{k+r}{r} \binom{k+s}{s} \binom{k}{n} \binom{k}{m}.$$

where, $F_Y(0) = 0$ and $F_X(0) = 0$,

then,

$$\Lambda(x,y) = \sum_{k=r=s=n=m=0}^{\infty} C \left[ \frac{(F_X(x))^{k+n+r+1} (F_Y(y))^{k+m+s+1}}{(k+n+r+1)(k+m+s+1)} + \lambda \left( \frac{2(F_X(x))^{k+n+r+2}}{k+n+r+2} - \frac{(F_X(x))^{k+n+r+1}}{k+n+r+1} \right) \left( \frac{2(F_Y(y))^{k+m+s+2}}{k+m+s+2} - \frac{(F_Y(y))^{k+m+s+1}}{k+m+s+1} \right) \right]$$

$$= \sum_{k=r=s=n=m=0}^{\infty} C \left[ \frac{\left(1 - e^{-\left(\alpha_1 x + \frac{\beta_1}{2} x^2\right)}\right)^{k+n+r+1} \left(1 - e^{-\left(\alpha_2 y + \frac{\beta_2}{2} y^2\right)}\right)^{k+m+s+1}}{(k+n+r+1)(k+m+s+1)} + \lambda \left( \frac{2\left(1 - e^{-\left(\alpha_1 x + \frac{\beta_1}{2} x^2\right)}\right)^{k+n+r+2}}{k+n+r+2} - \frac{\left(1 - e^{-\left(\alpha_1 x + \frac{\beta_1}{2} x^2\right)}\right)^{k+n+r+1}}{k+n+r+1} \right) \left( \frac{2\left(1 - e^{-\left(\alpha_2 y + \frac{\beta_2}{2} y^2\right)}\right)^{k+m+s+2}}{k+m+s+2} - \frac{\left(1 - e^{-\left(\alpha_2 y + \frac{\beta_2}{2} y^2\right)}\right)^{k+m+s+1}}{k+m+s+1} \right) \right]$$

(30)

## 4.2 Replacement

At each system failure, the conditional failure intensity unction returns to its value at time 0. Hence, if $n \geq 0$ and $X_n \leq x < X_{n+1}$, $Y_n \leq y < Y_{n+1}$, then

$$\lambda_c(x, y) = r(x - X_n, y - Y_n),$$

(31)

The counting process $\{N_2(x,y), x \geq 0, y \geq 0\}$ is a two-dimensional **RP** and the variables $(Z_n, T_n), n \geq 1$, are i.i.d. with *FGM* distribution function $F(x,y)$. It then follows that

$$P_n(x,y) = P[N_2(x,y) = n] = F^{(n)}(x,y) - F^{(n+1)}(x,y), n \geq 0$$

(32)

where $F^{(n)}(x,y)$ is the n-fold bivariate convolution of $F(x,y)$ with itself. The expected number of failures over $[0,x) \times [0,y)$ is given by

$$M(x,y) = E[N_2(x,y)] = \sum_{n=1}^{\infty} F^{(n)}(x,y),$$

(33)

which corresponds to the univariate form. As for the univariate function for *FGM*, the recursive statement of (33) is the renewal function "key integral renewal equation" and this can also be expressed as the solution of the two-dimensional integral equation

$$M(x,y) = F(x,y) + \int_0^x \int_0^y M(x-u, y-v) dv\, du.$$

(34)

and this function is the basis for analysis of the renewal process.

Equations (32)-(34) are from **Hunter** (1974), who also describes the bivariate Laplace transform approach for evaluating $M(x,y)$ using (33).

The Laplace (*Laplace–Stieltjes*) transform is the usual method of analysis for the renewal models. For the bivariate case, the Laplace transform of the *pdf* associated with the *Cdf*, $F(x,y)$ for *FGM* is:

$$f^*(s_1, s_2) = \int_0^\infty \int_0^\infty e^{-s_1 x - s_2 y} f(x,y) dy\, dx,$$

(35)

$$f^*(s_1,s_2) = \int_0^\infty \int_0^\infty e^{-s_1 x - s_2 y} f_X(x) f_Y(y)[1 + \lambda(2F_X(x) - 1)(2F_Y(y)-1)] dy\, dx,$$

$$= \int_0^\infty (\alpha_1 + \beta_1 x) e^{-\left((s_1+\alpha_1)x+\frac{\beta_1}{2}x^2\right)} \int_0^\infty (\alpha_2 + \beta_2 y) e^{-\left((s_2+\alpha_2)y+\frac{\beta_2}{2}y^2\right)} \Big[1$$

$$+ \lambda\left(1 - 2e^{-\left(\alpha_1 x+\frac{\beta_1}{2}x^2\right)}\right)\left(1 - 2e^{-\left(\alpha_2 y+\frac{\beta_2}{2}y^2\right)}\right)\Big] dy\, dx,$$

The series expansions of $e^{-\frac{\beta_1}{2}x^2}$ and $e^{-\frac{\beta_2}{2}y^2}$ are

$$e^{-\frac{\beta_1}{2}x^2} = \sum_{n=0}^\infty \frac{(-1)^n \beta_1^n}{2^n\, n!} x^{2n}$$

(36)

and

$$e^{-\frac{\beta_2}{2}y^2} = \sum_{m=0}^\infty \frac{(-1)^m \beta_2^m}{2^m\, m!} y^{2m}.$$

(37)

Then,

$$\int_0^\infty (\alpha_2 + \beta_2 y) e^{-\left((s_2+\alpha_2)y+\frac{1}{2}\beta_2 y^2\right)} dy$$

$$= \sum_{m=0}^\infty \frac{(-1)^m \beta_2^m\, (2m)!}{2^m\, m!\, (s_2+\alpha_2)^{2m}} \left[\frac{\alpha_2(s_2+\alpha_2) + \beta_2\,(2m+1)}{(s_2+\alpha_2)}\right],$$

(38)

and

$$\int_0^\infty (\alpha_2 + \beta_2 y)\, e^{-\left((s_2+\alpha_2)y+\frac{\beta_2}{2}y^2\right)} \left(1 - 2e^{-\left(\alpha_2 y+\frac{\beta_2}{2}y^2\right)}\right) dy$$

$$= \sum_{m=0}^{\infty} \frac{(-1)^m \beta_2^m (2m)!}{2^m m!} \left[\frac{\alpha_2 (s_2 + \alpha_2) + \beta_2 (2m+1)}{(s_2+\alpha_2)^{2m}}\right.$$

$$\left. - \frac{2^{m+1}(\alpha_2(s_2+2\alpha_2) + \beta_2(2m+1))}{(s_2+2\alpha_2)^{2m+1}}\right],$$

(39)

By similar for integral of $x$.

So, we obtained

$$f^*(s_1, s_2)$$
$$= \sum_{n=m=0}^{\infty} \frac{(-1)^{n+m} \beta_1^n \beta_2^m (2n)!\,(2m)!}{2^{n+m} m!\,n!\,(s_1+\alpha_1)^{2n} (s_2+\alpha_2)^{2m}} \left\{\left[\frac{(\alpha_1 s_1 + \alpha_1^2 + 2n\beta_1 + \beta_1)(\alpha_2 s_2 + \alpha_2^2 + 2m\beta_2 + \beta_2)}{(s_1+\alpha_1)(s_2+\alpha_2)}\right]\right.$$
$$+ \lambda\left[\frac{(\alpha_1 s_1 + \alpha_1^2 + 2n\beta_1 + \beta_1)}{(s_1+\alpha_1)^{2n}}\right.$$
$$\left. - \frac{2^{n+1}(\alpha_1 s_1 + 2\alpha_1^2 + 2n\beta_1 + \beta_1)}{(s_1+2\alpha_1)^{2n+1}}\right]\left[\frac{(\alpha_2 s_2 + \alpha_2^2 + 2m\beta_2 + \beta_2)}{(s_2+\alpha_2)^{2m}}\right.$$
$$\left.\left.- \frac{2^{m+1}(\alpha_2 s_2 + 2\alpha_2^2 + 2m\beta_2 + \beta_2)}{(s_2+2\alpha_2)^{2m+1}}\right]\right\}.$$

(40)

Using (40) in the analysis of the key renewal equation leads to

$$M^*(s_1, s_2) = \frac{F^*(s_1, s_2)}{1 - f^*(s_1, s_2)} = \frac{f^*(s_1, s_2)}{s_1 s_2 [1 - f^*(s_1, s_2)]}$$

So,

$$M^*(s_1, s_2)$$

$$= \left( \sum_{n=m=0}^{\infty} \frac{(-1)^{n+m} \beta_1^n \beta_2^m (2n)! (2m)!}{2^{n+m} m! n! (s_1 + \alpha_1)^{2n} (s_2 + \alpha_2)^{2m}} \left\{ \left[ \frac{(\alpha_1 s_1 + \alpha_1^2 + 2n\beta_1 + \beta_1)(\alpha_2 s_2 + \alpha_2^2 + 2m\beta_2 + \beta_2)}{(s_1 + \alpha_1)(s_2 + \alpha_2)} \right] \right. \right.$$

$$+ \lambda \left[ \frac{(\alpha_1 s_1 + \alpha_1^2 + 2n\beta_1 + \beta_1)}{(s_1 + \alpha_1)^{2n}} \right.$$

$$\left. - \frac{2^{n+1}(\alpha_1 s_1 + 2\alpha_1^2 + 2n\beta_1 + \beta_1)}{(s_1 + 2\alpha_1)^{2n+1}} \right] \left[ \frac{(\alpha_2 s_2 + \alpha_2^2 + 2m\beta_2 + \beta_2)}{(s_2 + \alpha_2)^{2m}} \right.$$

$$\left. \left. \left. - \frac{2^{m+1}(\alpha_2 s_2 + 2\alpha_2^2 + 2m\beta_2 + \beta_2)}{(s_2 + 2\alpha_2)^{2m+1}} \right] \right\} \right) \left( s_1 s_2 \left[ 1 \right. \right.$$

$$- \sum_{n=m=0}^{\infty} \frac{(-1)^{n+m} \beta_1^n \beta_2^m (2n)! (2m)!}{2^{n+m} m! n! (s_1 + \alpha_1)^{2n} (s_2 + \alpha_2)^{2m}} \left\{ \left[ \frac{(\alpha_1 s_1 + \alpha_1^2 + 2n\beta_1 + \beta_1)(\alpha_2 s_2 + \alpha_2^2 + 2m\beta_2 + \beta_2)}{(s_1 + \alpha_1)(s_2 + \alpha_2)} \right] \right.$$

$$+ \lambda \left[ \frac{(\alpha_1 s_1 + \alpha_1^2 + 2n\beta_1 + \beta_1)}{(s_1 + \alpha_1)^{2n}} \right.$$

$$\left. - \frac{2^{n+1}(\alpha_1 s_1 + 2\alpha_1^2 + 2n\beta_1 + \beta_1)}{(s_1 + 2\alpha_1)^{2n+1}} \right] \left[ \frac{(\alpha_2 s_2 + \alpha_2^2 + 2m\beta_2 + \beta_2)}{(s_2 + \alpha_2)^{2m}} \right.$$

$$\left. \left. \left. \left. - \frac{2^{m+1}(\alpha_2 s_2 + 2\alpha_2^2 + 2m\beta_2 + \beta_2)}{(s_2 + 2\alpha_2)^{2m+1}} \right] \right\} \right] \right)^{-1}.$$

(41)

Which correspond to the univariate forms.

## 5. Monotonicity of the (Reversed) Hazard Rate of the (Maximum) Minimum in Bivariate FGM Distributions

It is well known that in the case of independent random variables, the (reversed) hazard rate of the (maximum) minimum of two random variables is the sum of the individual (reversed) hazard rates and hence the monotonicity of the (reversed) hazard rate of the marginals is preserved by the monotonicity of the (reversed) hazard rate of the (maximum) minimum. However, for the bivariate distributions this property is not always preserved.

In case of the *FGM* family, we obtain the (reversed) hazard rate of the (maximum) minimum and provide several examples in some of which the (reversed) hazard rate is monotonic and in others it is non-monotonic. The *FGM* distributions of the (maximum) minimum of two random variables $x, y$ play an important role in various statistical applications. For example in the competing risks survival analysis due to two causes, $x$ and $y$ are not observed but $T_1 = Min(x, y)$ is the observable time of death. Similarly, in reliability studies, $T_1 = Min(x, y)$ is observed if the components are arranged in a series system, $T_2 = Max(x, y)$ is observed if the components are arranged in a parallel system.

The distribution function of $T_2$ for the $FGM$ is given by

$$F_{T_2}(t) = F_{X,Y}(t,t) = F_X(t) F_Y(t) \left[1 + \lambda(1 - F_X(t))(1 - F_Y(t))\right]$$

$$= \left(1 - e^{-(\alpha_1 t + \frac{1}{2}\beta_1 t^2)}\right)\left(1 - e^{-(\alpha_2 t + \frac{1}{2}\beta_2 t^2)}\right)\left[1 + \lambda e^{-((\alpha_1 + \alpha_2)t + \frac{1}{2}(\beta_1 + \beta_2)t^2)}\right].$$

(42)

The distribution function of $T_2$ for $FGM$ illustrated in Figure 3 for several dependence parameter values $\lambda$ and with parameter values $\alpha_1 = 0.5, \alpha_2 = 0.7, \beta_1 = 1.5$ and $\beta_2 = 2$.

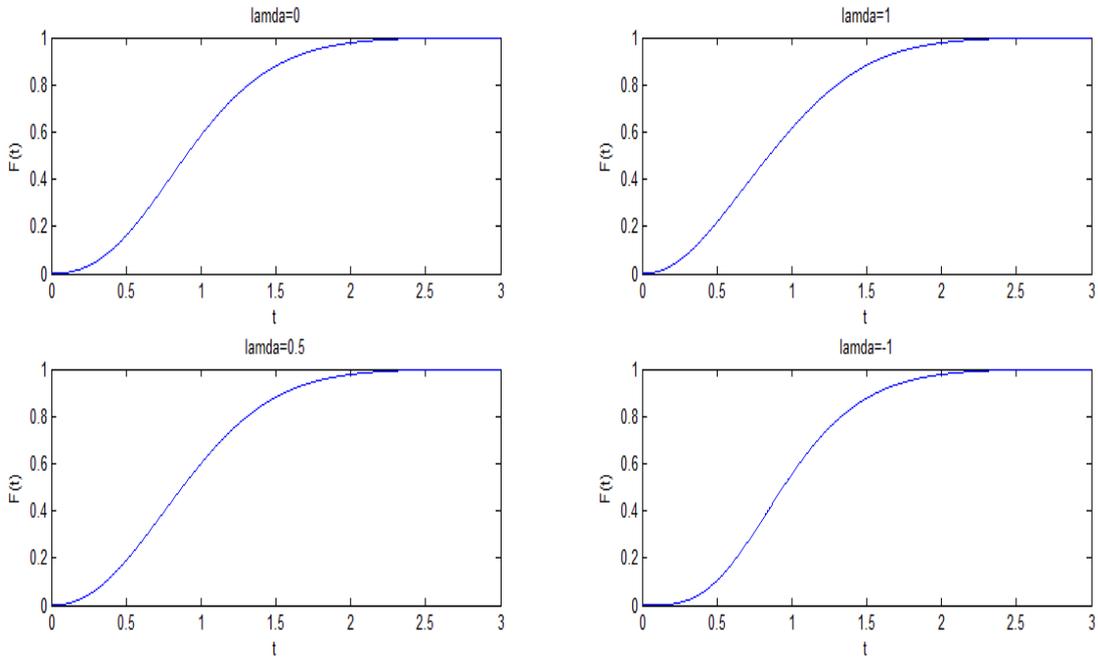

**Figure 3.** *Distribution Function of $T_2$ for the FGM for different dependence values of $\lambda$*

Thus the reversed hazard rate of $T_2$ is given by

$$rh_{T_2}(t) = \frac{d}{dt} \ln F_{T_2}(t),$$

$$rh_{T_2}(t) = rh_X(t) + rh_Y(t) - \frac{f_X(t)\big(1 - F_Y(t)\big) + f_Y(t)\big(1 - F_X(t)\big)}{1 + \lambda\big(1 - F_X(t)\big)\big(1 - F_Y(t)\big)}$$

(43)

Where $f_X(t)$ and $f_Y(t)$ are the marginal probability density functions (pdfs) and $rh_X(t)$ and $rh_Y(t)$ are the corresponding reversed hazard rates of $x$ and $y$ respectively.

$$f_X(t) = (\alpha_1 + \beta_1 t)e^{-(\alpha_1 t + \frac{1}{2}\beta_1 t^2)},$$

$$f_Y(t) = (\alpha_2 + \beta_2 t)e^{-(\alpha_2 t + \frac{1}{2}\beta_2 t^2)},$$

$$rh_X(t) = \frac{d}{dt}\ln f_X(t) = \frac{d}{dt}\left(\ln(\alpha_1 + \beta_1 t) - \left(\alpha_1 t + \frac{1}{2}\beta_1 t^2\right)\right) = \frac{\beta_1 - (\alpha_1 + \beta_1 t)^2}{(\alpha_1 + \beta_1 t)}$$

and

$$rh_Y(t) = \frac{d}{dt}\ln f_Y(t) = \frac{d}{dt}\left(\ln(\alpha_2 + \beta_2 t) - \left(\alpha_2 t + \frac{1}{2}\beta_2 t^2\right)\right) = \frac{\beta_2 - (\alpha_2 + \beta_2 t)^2}{(\alpha_2 + \beta_2 t)}.$$

then,

$$rh_{T_2}(t) = \frac{\beta_1 - (\alpha_1 + \beta_1 t)^2}{(\alpha_1 + \beta_1 t)} + \frac{\beta_2 - (\alpha_2 + \beta_2 t)^2}{(\alpha_2 + \beta_2 t)}$$
$$- \frac{\left((\alpha_1 + \alpha_2) + \frac{1}{2}(\beta_1 + \beta_2)t\right)e^{-\left((\alpha_1 + \alpha_2)t + \frac{1}{2}(\beta_1 + \beta_2)t^2\right)}}{\left[1 + \lambda e^{-\left((\alpha_1 + \alpha_2)t + \frac{1}{2}(\beta_1 + \beta_2)t^2\right)}\right]}.$$

(44)

The survival of $T_1$ for the *FGM* distribution is given by

$$\bar{F}_{T_1}(t) = \bar{F}(t,t) = \bar{F}_X(t)\bar{F}_Y(t)[1 + \lambda F_X(t)F_Y(t)]$$

(45)

Thus the hazard rate of $T_1$ is given by

$$h_{T_1}(t) = -\frac{d}{dt}\ln \bar{F}_{T_1}(t),$$

$$h_{T_1}(t) = h_X(t) + h_Y(t) - \lambda \left[ \frac{f_X(t)F_Y(t)+f_Y(t)F_X(t)}{1 + \lambda F_X(t)F_Y(t)} \right],$$

where $h_X(t)$ and $h_Y(t)$ are the hazard rates of X and Y respectively.

$$h_X(t) = -\frac{d}{dt}\ln(1 - F_X(t)) = \frac{d}{dt}\left(\alpha_1 t + \tfrac{1}{2}\beta_1 t^2\right) = (\alpha_1 + \beta_1 t)$$

and

$$h_Y(t) = -\frac{d}{dt}\ln(1 - F_Y(t)) = \frac{d}{dt}\left(\alpha_2 t + \tfrac{1}{2}\beta_2 t^2\right) = (\alpha_2 + \beta_2 t).$$

Then,

$$h_{T_1}(t) = \big((\alpha_1 + \alpha_2) + (\beta_1 + \beta_2)t\big)$$

$$- \lambda \left\{ \frac{(\alpha_1 + \beta_1 t)e^{-(\alpha_1 t+\tfrac{1}{2}\beta_1 t^2)}\left(1 - e^{-(\alpha_2 t+\tfrac{1}{2}\beta_2 t^2)}\right)}{\left[1 + \lambda\left(1 - e^{-(\alpha_1 t+\tfrac{1}{2}\beta_1 t^2)}\right)\left(1 - e^{-(\alpha_2 t+\tfrac{1}{2}\beta_2 t^2)}\right)\right]} \right.$$

$$\left. + \frac{(\alpha_2 + \beta_2 t)e^{-(\alpha_2 t+\tfrac{1}{2}\beta_2 t^2)}\left(1 - e^{-(\alpha_1 t+\tfrac{1}{2}\beta_1 t^2)}\right)}{\left[1 + \lambda\left(1 - e^{-(\alpha_1 t+\tfrac{1}{2}\beta_1 t^2)}\right)\left(1 - e^{-(\alpha_2 t+\tfrac{1}{2}\beta_2 t^2)}\right)\right]} \right\}.$$

(46)

Note that when λ=0, in the case of independence,

$$rh_{T_2}(t) = rh_X(t) + rh_Y(t)$$

and

$$h_{T_2}(t) = h_X(t) + h_Y(t).$$

## 6. Properties of Bivariate FGM Distribution with Proportional Hazard Rate Models

### 6.1 Dependence Measures

In order to study the dependence between the random variables $x$ and $y$ in *FGM* distribution, we consider the local dependence function, defined by (**Holland** and **Wang** (1987))

$$\gamma(x,y) = \frac{\partial^2}{\partial x\, \partial y} \ln f(x,y), \gamma(x,y)$$

$$= \frac{\partial^2}{\partial x\, \partial y} \Big[\ln(\alpha_2 + \beta_2 y) - (\alpha_2 y + \tfrac{1}{2}\beta_2 y^2) + \ln(\alpha_1 + \beta_1 x) - (\alpha_1 x + \tfrac{1}{2}\beta_1 x^2) +$$

$$\ln\left(1 + \lambda\left(1 - 2e^{-(\alpha_1 x + \tfrac{1}{2}\beta_1 x^2)}\right)\left(1 - 2e^{-(\alpha_2 y + \tfrac{1}{2}\beta_2 y^2)}\right)\right)\Big]$$

$$= 2\lambda(\alpha_2 + \beta_2 y)e^{-(\alpha_2 y + \tfrac{1}{2}\beta_2 y^2)}\Big[2(\alpha_1 + \beta_1 x)e^{-(\alpha_1 x + \tfrac{1}{2}\beta_1 x^2)}\Big(1$$

$$+ \lambda\left(1 - 2e^{-(\alpha_1 x + \tfrac{1}{2}\beta_1 x^2)}\right)\left(1 - 2e^{-(\alpha_2 y + \tfrac{1}{2}\beta_2 y^2)}\right)\Big)\Big]\Big[1$$

$$+ \lambda\left(1 - 2e^{-(\alpha_1 x + \tfrac{1}{2}\beta_1 x^2)}\right)\left(1 - 2e^{-(\alpha_2 y + \tfrac{1}{2}\beta_2 y^2)}\right)\Big]^{-2}.$$

(47)

The definition of total positive of order 2 ($TP_2$) functions and reverse rule of order 2 ($RR_2$) functions is the following.

***Definition1.*** APDF $f(x,y)$ is said to be $TP_2$ ($RR_2$) if

$$f(x,y)f(u,v) - f(x,v)f(u,y) \geq 0 \quad (\leq 0)$$

for all $x \leq u$ and $y \leq v$.

These properties are the strongest to fall dependence notions existing in the literature. Other dependence properties can be found in ***Joe*** (1997). The following result relates the local dependence function $\gamma(x,y)$ with the $TP_2$ and $RR_2$ properties (seeTheorem7, ***Holland*** and ***Wang*** (1987)).

***Theorem1.*** Let $f(x,y)$ be the PDF of $(X,Y)$ with support on a set S where the set S = $S_1 \times S_2$. Then $f(x,y)$ is $TP_2$ ($RR_2$) if and only if $\gamma(x,y) \geq 0$ ($\leq 0$).

Where $x, y \geq 0$, $\alpha_1, \alpha_2, \beta_1, \beta_2 \geq 0$ and $|\lambda| \leq 1$, then $\gamma(x,y) \geq 0$. So, $f(x,y)$ for *FGM* distribution is $TP_2$.

## 6.2 Hazard Gradient Functions

***Johnson*** and ***Kotz*** (1975) defined the hazard gradient as the vector

$$h(x,y) = \big(h_1(x,y), h_2(x,y)\big)^T = -\nabla \ln \bar{F}(x,y), \tag{48}$$

where $\nabla = (\frac{\partial}{\partial x}, \frac{\partial}{\partial y})^T$, $h_1(x,y)$ is the hazard rate of the conditional distribution of X given $(Y > y)$ and $h_2(x,y)$ is the hazard rate of the conditional distribution of Y given $(X > x)$, that is,

$$h_1(x,y) = h_{(X\backslash Y>y)}(x) = -\frac{\partial}{\partial x}\ln \bar{F}(x,y) = \frac{1}{\bar{F}(x,y)}\int_y^\infty f(x,v)dv$$

$$= \frac{f_X(x)}{\bar{F}(x,y)}\int_y^\infty f_V(v)[1 + \lambda(2F_X(x) - 1)(2F_V(v)-1)]dv$$

then,

$$h_1(x,y) = \frac{f_X(x)}{\bar{F}(x,y)}\left[(1 - F_Y(y)) + \lambda(2F_X(x) - 1)\left(\left(1 - (F_Y(y))^2\right) - (1 - F_Y(y))\right)\right]$$

$$= \left[(\alpha_1 + \beta_1 x)e^{-(\alpha_1 x + \frac{1}{2}\beta_1 x^2)}\right]\left[\left(e^{-(\alpha_1 x + \alpha_2 y + \frac{1}{2}(\beta_1 x^2 + \beta_2 y^2))}\right)\right.$$

$$+ \lambda\left(\left(2e^{-(\alpha_1 x + \frac{1}{2}\beta_1 x^2)} - e^{-2(\alpha_1 x + \frac{1}{2}\beta_1 x^2)}\right)\right.$$

$$- \left(e^{-(\alpha_1 x + \frac{1}{2}\beta_1 x^2)}\right)\right)\left(\left(2e^{-(\alpha_2 y + \frac{1}{2}\beta_2 y^2)} - e^{-2(\alpha_2 y + \frac{1}{2}\beta_2 y^2)}\right)\right.$$

$$\left.\left.- \left(e^{-(\alpha_2 y + \frac{1}{2}\beta_2 y^2)}\right)\right)\right]^{-1}\left[\left(e^{-(\alpha_2 y + \frac{1}{2}\beta_2 y^2)}\right) + \lambda(1\right.$$

$$- 2e^{-(\alpha_1 x + \frac{1}{2}\beta_1 x^2)})\left(\left(2e^{-(\alpha_2 y + \frac{1}{2}\beta_2 y^2)} - e^{-2(\alpha_2 y + \frac{1}{2}\beta_2 y^2)}\right)\right.$$

$$\left.\left.- \left(e^{-(\alpha_2 y + \frac{1}{2}\beta_2 y^2)}\right)\right)\right].$$

(49)

By similar,

$$h_2(x,y) = h_{(Y\backslash X>x)}(y) = -\frac{\partial}{\partial y}\ln \bar{F}(x,y) = \frac{1}{\bar{F}(x,y)}\int_0^\infty f(u,y)du$$

$$= \frac{f_Y(y)}{\bar{F}(x,y)}\left[(1 - F_X(x)) + \lambda(2F_Y(y)\right.$$

$$\left.- 1)\left(\left(1 - (F_X(x))^2\right) - (1 - F_X(x))\right)\right],$$

$$= \left[(\alpha_2 + \beta_2 y)e^{-(\alpha_2 y+\frac{1}{2}\beta_2 y^2)}\right]\left[\left(e^{-(\alpha_1 x+\alpha_2 y+\frac{1}{2}(\beta_1 x^2+\beta_2 y^2))}\right)\right.$$
$$+ \lambda\left(\left(2e^{-(\alpha_1 x+\frac{1}{2}\beta_1 x^2)} - e^{-2(\alpha_1 x+\frac{1}{2}\beta_1 x^2)}\right)\right.$$
$$- \left(e^{-(\alpha_1 x+\frac{1}{2}\beta_1 x^2)}\right)\left(\left(2e^{-(\alpha_2 y+\frac{1}{2}\beta_2 y^2)} - e^{-2(\alpha_2 y+\frac{1}{2}\beta_2 y^2)}\right)\right.$$
$$\left.\left.- \left(e^{-(\alpha_2 y+\frac{1}{2}\beta_2 y^2)}\right)\right)\right]^{-1}\left[\left(e^{-(\alpha_1 x+\frac{1}{2}\beta_1 x^2)}\right) + \lambda(1\right.$$
$$- 2e^{-(\alpha_2 y+\frac{1}{2}\beta_2 y^2)})\left(\left(2e^{-(\alpha_1 x+\frac{1}{2}\beta_1 x^2)} - e^{-2(\alpha_1 x+\frac{1}{2}\beta_1 x^2)}\right)\right.$$
$$\left.\left.- \left(e^{-(\alpha_1 x+\frac{1}{2}\beta_1 x^2)}\right)\right)\right].$$

(50)

**Lemma1.** If $f(x,y)$ is *TP₂* (*RR₂*), the conditional hazard rate $h_1(x,y)$ of X given $Y > y$ is decreasing (increasing) in y.

By using the above result and that *f* for *FGM* distribution is *TP₂*, then the conditional hazard rate $h_1(x,y)$ of X given $Y > y$ is decreasing.

## 6.3 The Clayton-Oakes Measure

In the context of bivariate survival models induced by frailties, **Oakes** (1989) considered the association measure

$$\theta(x,y) = \frac{\bar{F}(x,y)\, f(x,y)}{\bar{F}_1(x,y)\bar{F}_2(x,y)},$$

(51)

where,

$$\bar{F}_1(x,y) = \frac{\partial \bar{F}(x,y)}{\partial x},$$

$$= (F_Y(y) - 1)f_X(x) + \lambda f_X(x)(1 - 2F_X(x))\left(\left(1 - (F_Y(y))^2\right) - (1 - F_Y(y))\right)$$

$$= (-\alpha_1 - \beta_1 x)e^{-(\alpha_1 x+\alpha_2 y+\frac{1}{2}(\beta_1 x^2+\beta_2 y^2))} + \lambda(\alpha_1 + \beta_1 x)e^{-(\alpha_1 x+\frac{1}{2}\beta_1 x^2)}$$

$$\left(2e^{-(\alpha_1 x+\frac{1}{2}\beta_1 x^2)} - 1\right)\left(\left(2e^{-(\alpha_2 y+\frac{1}{2}\beta_2 y^2)} - e^{-2(\alpha_2 y+\frac{1}{2}\beta_2 y^2)}\right) - \left(e^{-(\alpha_2 y+\frac{1}{2}\beta_2 y^2)}\right)\right),$$

and

$$\bar{F}_2(x,y) = \frac{\partial \bar{F}(x,y)}{\partial y},$$

$$= (F_X(x) - 1)f_Y(y) + \lambda f_Y(y)(1 - 2F_Y(y))\left(\left(1 - (F_X(x))^2\right) - (1 - F_X(x))\right)$$

$$= (-\alpha_2 - \beta_2 y)e^{-(\alpha_1 x + \alpha_2 y + \frac{1}{2}(\beta_1 x^2 + \beta_2 y^2))} + \lambda(\alpha_2 + \beta_2 y)e^{-(\alpha_2 y + \frac{1}{2}\beta_2 y^2)}$$

$$\left(2e^{-(\alpha_2 y + \frac{1}{2}\beta_2 y^2)} - 1\right)\left(\left(2e^{-(\alpha_1 x + \frac{1}{2}\beta_1 x^2)} - e^{-2(\alpha_1 x + \frac{1}{2}\beta_1 x^2)}\right) - \left(e^{-(\alpha_1 x + \frac{1}{2}\beta_1 x^2)}\right)\right),$$

***Clayton*** (1978) obtained $\theta(x,y)$ deriving from the Cox *PHR* model, in a study of the association between the life spans of fathers and their sons.

Therefore, $\theta(x,y)$ for *FGM* distribution is

$$\theta(x,y)$$
$$= \left\{(\alpha_1 + \beta_1 x)(\alpha_2 + \beta_2 y)e^{-(\alpha_1 x + \alpha_2 y + \frac{1}{2}(\beta_1 x^2 + \beta_2 y^2))}\Big[1\right.$$
$$+ \lambda\left(1 - 2e^{-(\alpha_1 x + \frac{1}{2}\beta_1 x^2)}\right)\left(1 - 2e^{-(\alpha_2 y + \frac{1}{2}\beta_2 y^2)}\right)\Big]\right\}\left\{e^{-(\alpha_1 x + \alpha_2 y + \frac{1}{2}(\beta_1 x^2 + \beta_2 y^2))}\Big[1\right.$$
$$+ \lambda\left(1 - e^{-(\alpha_1 x + \frac{1}{2}\beta_1 x^2)}\right)\left(1 - e^{-(\alpha_2 y + \frac{1}{2}\beta_2 y^2)}\right)\Big]\right\}\left\{(-\alpha_1\right.$$
$$- \beta_1 x)e^{-(\alpha_1 x + \alpha_2 y + \frac{1}{2}(\beta_1 x^2 + \beta_2 y^2))}$$
$$+ \lambda(\alpha_1 + \beta_1 x)e^{-(\alpha_1 x + \frac{1}{2}\beta_1 x^2)}\left(2e^{-(\alpha_1 x + \frac{1}{2}\beta_1 x^2)}\right.$$
$$- 1\right)\left(\left(2e^{-(\alpha_2 y + \frac{1}{2}\beta_2 y^2)} - e^{-2(\alpha_2 y + \frac{1}{2}\beta_2 y^2)}\right) - \left(e^{-(\alpha_2 y + \frac{1}{2}\beta_2 y^2)}\right)\right)\right\}^{-1}\left\{(-\alpha_2\right.$$
$$- \beta_2 y)e^{-(\alpha_1 x + \alpha_2 y + \frac{1}{2}(\beta_1 x^2 + \beta_2 y^2))}$$
$$+ \lambda(\alpha_2 + \beta_2 y)e^{-(\alpha_2 y + \frac{1}{2}\beta_2 y^2)}\left(2e^{-(\alpha_2 y + \frac{1}{2}\beta_2 y^2)}\right.$$
$$- 1\right)\left(\left(2e^{-(\alpha_1 x + \frac{1}{2}\beta_1 x^2)} - e^{-2(\alpha_1 x + \frac{1}{2}\beta_1 x^2)}\right)\right.$$
$$- \left(e^{-(\alpha_1 x + \frac{1}{2}\beta_1 x^2)}\right)\right)\right\}^{-1}.$$

(52)

## 7. Conclusion

The Farlie-Gumbel-Morgenstern (*FGM*) family of bivariate linear exponential distributions given in this study. We discussed some statistical properties of the bivariate distribution, including reliability, MTTF, hazard function, minimal repair, replacement, monotonicity of the (reversed) hazard rate and proportional hazard rate models in this paper. Some figures are used to illustrate how the results obtained can be applied.